\documentclass{appolb}
\usepackage{amsmath}
\usepackage{amssymb} 
\usepackage[small]{caption2} 
\usepackage{fleqn} 
\usepackage{graphicx} 
\usepackage{mathrsfs}	
\usepackage[small,loose]{subfigure}  
\usepackage{cite} 
\usepackage{cancel}
\allowdisplaybreaks
\newif\ifappendix

\appendixfalse
\advance \headheight by 3.0truept       
\DeclareMathOperator{\re}{Re}
\DeclareMathOperator{\im}{Im}

\newcommand{\CenterObject}[1]{\ensuremath{\vcenter{\hbox{#1}}}}

\begin{document}
\preprint{}
\title{\appHuge{ De Sitter Vacua from Matter Superpotentials }}
\author{Oleg Lebedev,
Hans Peter Nilles,
Michael Ratz
\address{Physikalisches Institut der Universit\"at Bonn, Nussallee 12, 
53115 Bonn, Germany}
}
\maketitle
\begin{abstract}
Consistent uplifting of AdS vacua in string theory often requires extra light
degrees of freedom in addition to those of a (K\"ahler) modulus. Here we
consider the possibility that de Sitter and Minkowski vacua arise due to hidden
sector matter interactions. We find that, in this scheme, the  hierarchically
small supersymmetry breaking scale can be  explained by the scale of gaugino
condensation and that interesting patterns of the soft terms arise. In
particular, a matter--dominated supersymmetry breaking scenario and a version of
the mirage mediation scheme appear in the framework of spontaneously broken
supergravity.
\end{abstract}
  
\section{Introduction}

Fluxes on an internal manifold allow one to stabilize most 
moduli \cite{Giddings:2001yu}, 
but usually not all. In particular, in the KKLT model \cite{Kachru:2003aw},
the overall K\"ahler modulus $T$  is not fixed by the fluxes and is stabilized
by non--perturbative effects such as gaugino condensation \cite{Nilles:1982ik}. 
The corresponding superpotential 
\begin{equation}
 W~=~ W_0 + A {\rm e}^{-a T}
\end{equation}
leads to an AdS supersymmetric minimum. To obtain a realistic vacuum, this
minimum has to be uplifted. The original KKLT proposal was to use an explicit SUSY 
breaking term induced by anti D3 branes,
\begin{equation}
\Delta V~=~ {k \over (T + \overline{T})^2} \;,
\end{equation}
to do the uplifting. A  somewhat  more appealing possibility is to employ  the 
supersymmetric D--terms for this purpose \cite{Burgess:2003ic},
\begin{equation}
\Delta V~=~ {1 \over 2 g^2} D^2 \;.
\end{equation}
However, a supersymmetric minimum cannot be uplifted by the D--terms 
\cite{Choi:2005ge}. It is  possible to uplift non--supersymmetric minima
which arise once $\alpha'$ corrections  \cite{Becker:2002nn}   have been included 
\cite{Balasubramanian:2005zx}.
In any case, this procedure relies on the presence of charged matter 
in the effective theory   \cite{Achucarro:2006zf}. 
Thus, it appears that the uplifting within the supergravity framework requires
extra degrees of freedom in addition to those of a K\"ahler modulus. 
This perhaps is not always the case, but at least it is true for simple
K\"ahler potentials.
Then one may ask whether it is necessary to use the D--terms at all: 
de Sitter vacua may simply result from  the  superpotential    interactions
with the extra degrees of freedom. We note also that in models with the D--term uplifting
it would be very difficult to obtain a hierarchically small SUSY breaking scale
\cite{Achucarro:2006zf},\cite{Villadoro:2005yq}.\footnote{We thank K.~Choi for
pointing out generic difficulties of D--term uplifting with
hierarchically small gravitino mass.}

In this work, we study the possibility that dS and Minkowski vacua arise
due to interactions of hidden matter. We identify the local superpotential
structures realizing this situation and study the resulting soft SUSY breaking terms.
We find that interesting patterns arise. In particular, a matter--dominated
SUSY breaking scenario and a version of the mirage mediation scheme 
appear in the context of spontaneously broken supergravity.

\section{Minkowski and de Sitter vacua due to matter interactions}

Let us start by reviewing the supergravity formalism.
The supergravity scalar potential is expressed in terms of the function 
\begin{equation}
 G~=~K + \ln \vert W \vert^2 \;,
\end{equation}
with $K$ and $W$ being the K\"ahler potential and the superpotential, respectively,
as
\begin{equation}
V~=~ {\rm e}^{G} \bigl(   G_i G_{\bar \jmath} G^{i \bar \jmath}  -3   \bigr) \;.
\end{equation}
Here the subscript $i$ denotes differentiation with respect to $i$-th field  
and $G^{i \bar \jmath} $ is the inverse  K\"ahler metric. The SUSY breaking F--terms are 
found from 
\begin{equation}
F^m ~=~ {\rm e}^{G/2} G^{m \bar n} G_{\bar n}
\end{equation}
evaluated at the minimum of the potential. The gravitino mass is
\begin{equation}
m_{3/2}~=~ {\rm e}^{G/2} \;.
\end{equation}

In what follows, we study under which circumstances de Sitter and Minkowski vacua
arise in supergravity models involving a modulus ($T$) and a matter field ($C$).

\subsection{No go with a single modulus}

In this subsection, we show that de Sitter  vacua 
are not possible in models with a single modulus   as long as the K\"ahler potential
takes on its classical form 
\begin{equation}
K~=~-a \ln (T+\overline{T}) \;.
\end{equation}
Here $1\le a \le3$ depending on the nature of the modulus.

The scalar potential reads
\begin{equation}
V~=~ {1 \over (T+ \overline{T})^a} ~\biggl(
{1\over a} \Bigl\vert  W_T (T+ \overline{T}) 
-a\, W\,  \Bigr\vert^2 -3 \vert W\vert^2
\biggr) \;.
\end{equation}
The stationary point condition  $\partial V / \partial T=0$   is then
\begin{eqnarray}
&&(\overline{W}_T (T+ \overline{T}) -a \overline{W}) \,
\Bigl( W_{TT}(T +\overline{T}) +(1-a) W_T \Bigr) {T+\overline{T} \over a}
-  \nonumber\\
&&  (W_T (T+ \overline{T}) -a W) \Bigl(
\overline{W}_T (T+ \overline{T}) {a-1 \over a} + \overline{W} (3-a)
\Bigr)=0 \;.
\end{eqnarray}
To analyze stability of the stationary point, we need the second derivatives of
the potential. Using the above equation, one can write $\partial^2 V / \partial
T  \partial \overline{T}$ in a compact form,
\begin{equation}
{ \partial^2 V \over  \partial T \, \partial \overline{T}  } ~=~
- {2\over (T+ \overline{T})^2} \Bigl(     V_0 +{3-a \over (T+\overline{T})^a } \vert W \vert^2      \Bigr)
 \;.
\end{equation}
For  $a\leq 3$ and $V_0 \geq 0$  , this expression is non--positive  which implies
that at least one of the eigenvalues of the Hessian is negative or zero.
Thus  realistic   dS/Minkowski  minima are not possible.
This result was also found numerically in \cite{Brustein:2004xn} (see also \cite{scr}).

Our conclusion relies on the classical form of the K\"ahler potential. 
In particular, perturbative  $\alpha'$  corrections to the  K\"ahler potential
allow for dS vacua   \cite{Westphal:2005yz}. 
Also,  the separation
of the $G$--function into the K\"ahler potential and the superpotential is ambiguous.
With a fixed K\"ahler potential, integrating out heavy fields may lead to effects
which cannot be described by a holomorphic 
superpotential   \cite{deAlwis:2005tf}.
In this work, we will $assume$ that these effects are subdominant  in the region 
of interest. Then extra degrees of freedom are required to obtain dS vacua.
In what follows, we  will study the case when these additional degrees of freedom
are provided by  matter fields and analyze 
the local superpotential structure  allowing
for dS/Minkowski vacua.

\subsection{A modulus and a matter field}

Suppose that the low energy theory involves a modulus $T$   and a matter field $C$.
The corresponding  K\"ahler potential is
\begin{equation}
 K~=~-3 \ln (T+\overline{T}) + \vert C \vert^2 \;,
\end{equation}
where we have assumed for definiteness that $T$ is an overall K\"ahler modulus
and $C$ has an effective ``modular weight'' zero. Systems of this type arise
in type IIB and  heterotic string theory. The effective superpotential  obtained
by integrating out heavy moduli and matter fields is assumed to be of the form
\begin{equation}
 W~=~ \sum_i \omega_i(C)~ {\rm e}^{-\alpha_i T}  + \phi (C) \;,
\label{w}
\end{equation}
where the sum runs over gaugino condensates \cite{Krasnikov:1987jj}. The functions $\omega_i(C)$ and 
$\phi (C)$  arise due to perturbative and non--perturbative interactions
in the process of integrating out heavy fields. We will treat them as some generic functions
since only their local behaviour is important for our purposes. In particular,
we will allow for linear terms $ \propto C $ which can arise from interactions
with heavy matter fields $s_i$  , $\Delta W \sim C \langle s_1...s_N   \rangle {\rm e}^{-\alpha T} $.
We assume that $C$ is a singlet under unbroken gauge symmetries.

The supergravity scalar potential is given by
\begin{equation}
V~=~ {{\rm e}^{C\, \overline{C}} \over (T + \overline{T})^3} \biggl[
{1\over 3} \Bigl\vert  W_T (T+ \overline{T}) -3 W  \Bigr\vert^2    
+  \vert W_C + W \overline{C} \vert^2                 -3 \vert W\vert^2
\biggr]   \;.
\end{equation}
It is convenient to introduce 
\begin{eqnarray}
&&f^T \equiv \overline{W}_T (T + \overline{T}) - 3 \overline{W} \;, \nonumber\\
&&f^C \equiv \overline{W}_C + \overline{W}\, C \;,
\end{eqnarray}
such that 
\begin{eqnarray}
&&F^T= {T+ \overline{T} \over 3 \overline{W}  }~ m_{3/2}~f^T \;, \nonumber\\
&&F^C= {1 \over  \overline{W}  }~ m_{3/2}~f^C \;.
\label{F-f}
\end{eqnarray}
Then the stationary point conditions read
\begin{eqnarray}
 {\partial V \over \partial C}&=& V \overline{C} 
 +{{\rm e}^{C\,\overline{C}} \over (T + \overline{T})^3} \biggl[
{1\over 3} ( W_{TC} (T + \overline{T})-3 W_C   ) f^T 
+ (W_{CC} + W_C\, \overline{C}) f^C \nonumber\\ 
&&{}+
\overline{W} \bar f^C  -3 W_C \overline{W} \biggr]=0 \;, \nonumber\\
{\partial V \over \partial T}&=& - {3\over T+ \overline{T}} V 
+ {{\rm e}^{C\,\overline{C}} \over (T + \overline{T})^3} 
\biggl[ {1\over 3} (W_{TT} (T +\overline{T}) -2 W_T) f^T 
+ {1\over 3} \overline{W}_T \bar f^T 
\nonumber\\
&&{}+
(W_{TC}+ W_T\,\overline{C}) f^C - 3 W_T \overline{W} \biggr]=0 \;.
\label{dV=0}
\end{eqnarray}

We are interested in  local behaviour of the scalar potential. Without loss of generality,
assume that the above equations are satisfied at 
\begin{equation}
C=0 \;,~~~T=T_0 \;,
\end{equation}
then Eq.~(\ref{dV=0}) translates into relations among the derivatives of the superpotential
 at  that point. For the analysis of local  behaviour of the scalar potential, we only
need derivatives of the superpotential up to order three. Then $W$ can be written as
\begin{eqnarray}
W&=&W_0 + W_C C + W_T (T-T_0) +{1\over 2} W_{CC} C^2 + W_{TC} C (T-T_0) 
\nonumber\\
& & {}
+{1\over 2}W_{TT} (T-T_0)^2
+{1\over 6} W_{CCC} C^3 + {1\over 2} W_{TCC} C^2 (T-T_0)
\nonumber\\
 & &{}
 +   {1\over 2} W_{TTC} C (T-T_0)^2
+{1\over 6} W_{TTT}  (T-T_0)^3\;.    
\end{eqnarray}
Given vacuum energy $V_0$ and supersymmetry breaking parameters 
$f^T, f^C$  which measure the balance
between modulus and matter SUSY breaking as input, 
Eq.~(\ref{dV=0})   identifies  local superpotentials realizing this situation.
Stability considerations impose further constraints
 on the superpotential structure (see \cite{Gomez-Reino:2006dk},\cite{scr} 
on the related discussion). 

The superpotential expansion parameters can be expressed in terms of $F^T, F^C, V_0$
or, using Eq.~(\ref{F-f}), in terms of  $f^T, f^C, V_0$ (up to an irrelevant  phase)
as
\begin{eqnarray} 
\label{stat}
\vert W_0 \vert &=& {1\over \sqrt{3} } \biggl( 
{1\over3 } \vert f^T\vert^2 + \vert f^C \vert^2 -V_0 (T_0 + \overline{T}_0)^3
\biggr)^{1/2} \;, \\
W_C &=& \bar f^C \;, \nonumber\\
W_T &=& {3 W_0 + \bar f^T \over T_0 + \overline{T}_0} \;,\nonumber\\
W_{CC} &=& -{1\over 3} \biggl( W_{TC} (T_0 + \overline{T}_0 ) -3 \bar f^C \biggr) {f^T \over f^C} 
+ 2\overline{W}_0
  {\bar f^C \over f^C} \;, \nonumber\\
W_{TT} &=& {3\over (T_0 + \overline{T}_0) f^T} \biggl(
3 (T_0 + \overline{T}_0)^2 V_0 +{2\over 3} W_T f^T 
 -{1\over 3} \overline{W}_T \bar f^T -W_{TC} f^C 
+ 3 W_T \overline{W}_0
\biggr). \nonumber
\end{eqnarray}
Here the phase of $W_0$ is a free parameter. Also, 
$W_{TC}$ is a free parameter as long as $f^T \not=0$. If 
 $f^T =0$, $W_{TT}$ becomes a free parameter and $W_{TC}$ is found from
\begin{equation}
W_{TC}~=~ {3\over f^C} \Bigl(    
 (T_0 + \overline{T}_0)^2 V_0 + W_T \overline{W}_0
\Bigr) \;.
\end{equation}
In this work, we will only consider the case  $f^C \not=0$. 

Higher derivatives of the superpotential remain undetermined at this stage.
They are constrained by stability considerations.
To analyze stability of the
stationary point, one can neglect the vacuum energy, $V_0 \lll 1$, and use
 the following second derivatives of the potential
\begin{eqnarray}
 (T_0 + \overline{T}_0)^3 V_{C \overline{C}}&=& 
 {1\over 3} \vert W_{TC}(T_0 + \overline{T}_0)-3W_C\vert^2+
\vert W_{CC} \vert^2 + \vert W_0 \vert^2 - \vert f^C \vert^2 \;, \nonumber\\
 (T_0 + \overline{T}_0)^3 V_{C  C}
 &=& {1\over 3} (W_{TCC}(T_0 + \overline{T}_0)-3 W_{CC} ) f^T
+ W_{CCC} f^C - W_{CC} \overline{W}_0 \;, \nonumber\\
(T_0 + \overline{T}_0)^3 V_{T \overline{T}}
 &=&{1\over 3} \vert W_{TT}(T_0 + \overline{T}_0)-2W_T\vert^2+
\vert W_{TC} \vert^2 -{8\over 3}\vert W_{T} \vert^2\nonumber\\*
& & {} + \left( 
{1\over 3} W_{TT} f^T + {\rm h.c.} \right) \;, \nonumber\\
(T_0 + \overline{T}_0)^3 V_{T  T}&=& 
{1\over 3} (W_{TTT} (T_0 + \overline{T}_0) -W_{TT} ) f^T
+{2\over 3} (W_{TT}(T_0 + \overline{T}_0) -2 W_T  ) \overline{W}_T \nonumber\\
&&{}+ W_{TTC} \overline{W}_C -3 W_{TT} \overline{W}_0 \;, \nonumber\\
(T_0 + \overline{T}_0)^3 V_{T \overline{C}}
&=& {1\over 3} (W_{TT}(T_0 + \overline{T}_0) -2 W_T  )
(\overline{W}_{TC}(T_0 + \overline{T}_0)-3 \overline{W}_C ) 
+ {1\over 3} \overline{W}_{TC} \bar f^T
\nonumber\\
&&{}-  2 W_T f^C + W_{TC} \overline{W}_{CC} \;, \nonumber\\  
(T_0 + \overline{T}_0)^3 V_{T  C}
&=& {1\over 3} (W_{TTC}(T_0 + \overline{T}_0)-2 W_{TC} ) f^T
+{1\over 3} (W_{TC}(T_0 + \overline{T}_0)-3 W_{C} ) 
\overline{W}_T \nonumber\\
&&{}+   W_{TCC}  f^C -2 W_{TC} \overline{W}_0 \;.
\end{eqnarray}
The eigenvalues of $\partial^2 V / \partial x_i \partial \bar x_j$ must be positive.
This constrains $W_{TC}$ and higher derivatives of the superpotential.
The general formulae are unilluminating, so let us focus on the cases of interest,
in particular, matter--dominated SUSY breaking: 
$ 0  \le \vert f^T \vert \ll \vert f^C \vert $.

Consider the limit $\vert f^T \vert \ll \vert f^C \vert $. 
From Eq.~(\ref{stat}), this corresponds to large $W_{TT}$. 
Then  $\vert  V_{T \overline{T}} \vert \gg \vert  V_{C \overline{C}} \vert , 
\vert  V_{T \overline{C}} \vert $. To obtain a particularly simple
structure of $\partial^2 V / \partial x_i \partial \bar x_j$, let us choose
(otherwise unconstrained) $W_{CCC}, W_{TCC}, W_{TTC}$ such that the matrix 
elements $V_{TT}, V_{CC}, V_{TC}$ are small. Then we have 
\begin{equation}
{\partial^2 V \over \partial x_i \partial \bar x_j} ~\simeq~
{1\over (T_0 + \overline{T}_0)^3 }~ \left(
\begin{matrix}  
\vert A \vert^2 & 0 & A a & 0 \\ 
          0 &  \vert A \vert^2 & 0 & A^* a^* \\
         A^* a^* & 0 & \vert a \vert^2 +\Delta & 0\\
          0 & Aa & 0 & \vert a \vert^2 +\Delta  
\end{matrix}
\right) \;,
\end{equation}
where 
\begin{eqnarray}
 A &\equiv& {1\over \sqrt{3} } \overline{W}_{TT} (T_0 + \overline{T}_0) \;, \nonumber\\
 a &\equiv&  {1\over \sqrt{3} } (W_{TC}(T_0 + \overline{T}_0)-3 W_{C} ) \;, \nonumber\\
 \Delta &\equiv&  2 \vert W_0 \vert^2 \;, 
\end{eqnarray}
so that $\vert A \vert \gg \vert a \vert, \Delta$.
The order of the indices is defined by $(x_1, x_2, x_3, x_4)= 
(\overline{T}, T, \overline{C}, C)$.
All of the subdeterminants of this matrix are positive, hence the eigenvalues are positive.
This proves that the stationary point is a local minimum.

In  the case $f^T=0$,
$W_{TT}$ is a free parameter and can be taken to be large. Then the same argument applies.
In both cases, the spectrum consists of 2 heavy states with masses of order $\vert W_{TT}\vert$ 
and 2 lighter states with masses of order $\vert W_0 \vert$.

The above  local structure can be translated into constraints on the parameters of the 
original superpotential (\ref{w}). We note that large
$\vert W_{TT}\vert$ arises naturally in racetrack models since differentiation
by $T$ brings down  the factor  1/(beta function) and the moduli are 
heavy compared to $m_{3/2}$ (see
e.g.~\cite{Krasnikov:1987jj},\cite{Buchmuller:2004xr}),
\begin{equation}
 |W_{TT}|~\sim~\alpha^2\,|W_0|~\gg~|W_0|
\end{equation}
with $\alpha\sim1/(\text{beta function})$.
This means that $T$ is stabilized close to a supersymmetric point since
$\partial V/\partial T=0$ (cf.\ Eq.~(\ref{dV=0})) implies
\begin{equation}
 W_{TT}\,F^T+\text{smaller terms}~=~0\;,
\end{equation}
such that $F^T\sim m_{3/2}^2/|W_{TT}|\sim m_{3/2}/\alpha^2$.
Further, the scale of SUSY breaking is explained by the scale of gaugino
condensation, as long as  $\phi(C)$ is negligible.

\begin{figure}[!h]
\centerline{
\subfigure[$V(C,T_0)$.]{\CenterObject{\includegraphics[scale=0.9]{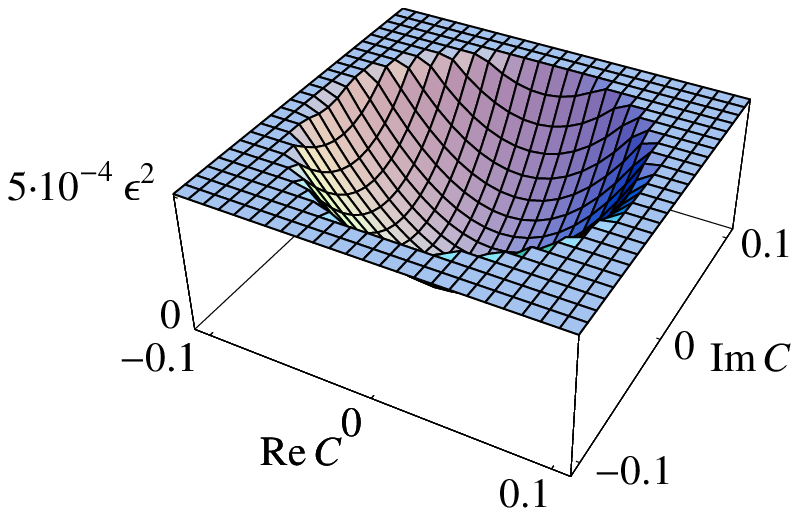}}}
\quad
\subfigure[$V(0,T)$.]{\CenterObject{\includegraphics[scale=0.9]{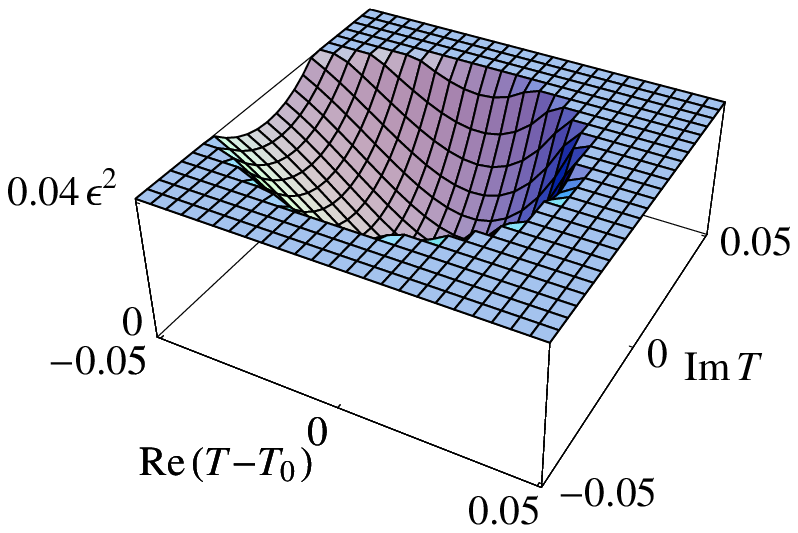}}}}
\caption{
The scalar potential in (a) $\re C$, $\im C$ and (b) $\re T$, $\im T$
 coordinates. The minimum is at $T=2$, $C=0$.
}
\label{potential1}
\end{figure}

{\bf Numerical example.}  If the observable matter is placed on D7 branes in
type IIB constructions, the K\"ahler modulus $T$ should be stabilized at $\re
T_0=2$ as required by the observed gauge couplings.  Consider now an  example
$f^C=\epsilon, f^T=0.03 \epsilon, V_0 \simeq 0$ for small  $\epsilon$ generated
by gaugino condensation.  An example of the local  superpotential structure 
realizing this situation is given by 
\begin{eqnarray}
W&\simeq& \epsilon\, \Bigl[  0.577 + C +  0.441\, (T-2) 
+  0.592\, C^2 + 9.595\, (T-2)^2+ 0.114\, C^3 \nonumber\\
&&\hphantom{\epsilon\, \Bigl[}{}+ 0.220\, C^2\, (T-2) + 46.451\, (T-2)^3  \Bigr]  \;.
\end{eqnarray}
The shape of the potential around the minimum is shown in Fig.~\ref{potential1}.


Similarly, one can construct examples with minima at $T_0 \sim 100$, where the
supergravity approximation is trustable. 

The main lesson here is that, unlike in the case of a single modulus, 
 dS/Minkowski vacua with interesting SUSY breaking patterns 
can be realized  in the framework of spontaneously broken supergravity.

\section{Patterns of the soft masses}

Let us now study the emerging patterns of the observable matter
soft terms. The scale of the soft terms is set by the gravitino
mass 
\begin{equation} 
 m_{3/2} ~=~ {\vert W_0 \vert \over (T_0 + \overline{T}_0)^{3/2}} \;,
\end{equation}
which is in turn generated via gaugino condensation, 
$ \vert W_0 \vert \sim \langle   \sum_i  {\rm e}^{-\alpha_i T}    \rangle $,
as long as the matter superpotential  $\phi (C)$  is negligible.  
The tree level soft terms are found from the general formulae,
\begin{eqnarray}
 M_a & = &{1\over 2} (\re f_a)^{-1} F^m \partial_m f_a \;, \nonumber\\
 m_\alpha^2 & = & m_{3/2}^2- \overline{F}^{\bar m} F^n \partial_{\bar m} \partial_n \ln K_\alpha
   \;, \nonumber\\
 A_{\alpha\beta\gamma} & = &F^m \Bigl[
\hat K_m + \partial_m \ln Y_{\alpha\beta\gamma} -\partial_m \ln (K_\alpha K_\beta K_\gamma)
\Bigr] \;,
\end{eqnarray}
where $m$ runs over SUSY breaking fields, $f_a$ are the gauge kinetic functions,
$K_\alpha$ is the K\"ahler metric for the  observable sector fields and $\hat
K_m \equiv \partial_m \hat K$ with $\hat K$ being the K\"ahler potential for the
hidden sector fields. The $\mu$ and $B\mu$ terms are not listed as their
generation mechanism is strongly model--dependent. These formulae are to be
amended by loop--suppressed terms such as the anomaly mediated contributions
\cite{Giudice:1998xp}.

The gauge kinetic functions are model dependent quantities. Consider, for definiteness,
type IIB string theory. For gauge fields on D7 branes, we have
\begin{equation} 
f_a~=~T \;,
\end{equation}
while in the case of D3 branes 
\begin{equation} 
f_a~=~ {\rm const} \;.
\end{equation}
The total K\"ahler potential is given by
\begin{equation}
 K~=~-3 \ln (T+ \overline{T}) + C\, \overline{C} 
 + Q_i \overline{Q}_i (T+ \overline{T})^{n_i} \Bigl[
 1+ \xi_i\, C\, \overline{C} + {\cal O} (C^4)
\Bigr] \;,
\end{equation}
where $Q_i$ are the observable fields with  ``effective modular weights'' $n_i$.
Here we include for generality quartic couplings between observable and hidden
sector fields, which can be present at tree level or generated radiatively
(see e.g. \cite{Choi:1997de}).

The resulting soft terms are 
\begin{eqnarray}
 M_a &=&  (0~{\rm or} ~1)\times  {F^T \over T_0 + \overline{T}_0} + {\rm anomaly}
 \;, \nonumber\\
 m_\alpha^2 &=& m_{3/2}^2 + n_\alpha {\vert F^T \vert^2 \over (T_0 +
 \overline{T}_0)^2 } 
- \xi_\alpha \vert F^C \vert^2   +  {\rm anomaly}  \;, \nonumber\\
 A_{\alpha\beta\gamma} & = &  -{F^T \over T_0 + \overline{T}_0} \left[
3+ n_\alpha + n_\beta + n_\gamma \right] +  {\rm anomaly} \;,
\end{eqnarray}
where we have assumed that $ Y_{\alpha\beta\gamma}$ are independent of $T$ and
$C$. The ``anomaly''  contributions generally include various loop--suppressed 
terms  (in addition to those due to the super-Weyl anomaly) which result from
regularization of the effective SUGRA 
\cite{Bagger:1999rd},\cite{Binetruy:2000md} and string threshold corrections
\cite{Dixon:1990pc}.
$F^T$ and $F^C$ are subject to the constraint
\begin{equation}
m_{3/2}^2~=~ {\vert F^T \vert^2 \over (T_0 + \overline{T}_0)^2 } +{1\over 3} \vert F^C \vert^2 \;.
\end{equation}
Below we consider two most interesting special cases: matter domination and mirage mediation.  These arise when the T--modulus is heavy,
\begin{equation}
\vert W_{TT} \vert~\gg~\vert W_0 \vert, \vert W_{T} \vert, \vert W_{CC} \vert,
\dots
\end{equation}
such that $T$ is stabilized close to a supersymmetric point. This situation is 
rather natural for gaugino condensation models due to the smallness of the beta 
functions of  condensing gauge groups  (see e.g. \cite{Krasnikov:1987jj},\cite{Buchmuller:2004xr}).

{\bf 1. Matter dominated SUSY breaking.}
This corresponds to $F^T=0$ such that
\begin{eqnarray}
&& M_a =   {\rm anomaly}
 \;, \nonumber\\
&&  m_\alpha^2 = m_{3/2}^2  (1-3 \xi_\alpha)   
  +  {\rm anomaly}  \;, \nonumber\\
&& A_{\alpha\beta\gamma}=    {\rm anomaly} \;.
\end{eqnarray}
A particularly simple case is $\xi_\alpha \sim 0$.
This provides an interesting ``regularization''  of the traditional 
anomaly mediation scheme
in the sense that it inherits main features of the latter while avoiding
tachyonic sfermions. We note that this scenario 
is different from the moduli--dominated models in the heterotic string in two aspects. 
First, the cosmological constant
here can be made arbitrarily small and positive. Second, the string threshold
corrections to the gauge kinetic functions are independent of $C$ (or, at least, negligible
at $C=0$) and the K\"ahler anomalies  \cite{Bagger:1999rd},\cite{Binetruy:2000md}
do not contribute to the gaugino masses.
Therefore,  $M_a$ receive a leading contribution from the super-Weyl anomaly,
as in the original version of anomaly mediation \cite{Giudice:1998xp}.

The soft terms exhibit the following hierarchy
\begin{equation}
M_a ~,~ A ~\ll ~m_{\rm scalar }~,~ m_{3/2} ~~,
\end{equation}
while for the T--modulus and the hidden matter we have $m_T \gg  m_{3/2}$ and
$m_C \sim  m_{3/2}$.

A solid feature of this SUSY breaking scenario is that the LSP is predominantly 
a wino and the mass splitting between the chargino and the neutralino is small.
This leads to spectacular collider signatures such as long lived charged
particle tracks \cite{Feng:1999fu}.

{\bf 2. Mirage mediation.} This scenario appears in the case 
$F^T / (T_0 + \overline{T}_0  ) \sim F^C / 4 \pi^2$ \cite{Choi:2005ge},\cite{Choi:2005uz}.
The modulus and the anomaly contribute to the gaugino masses and the A-terms 
in comparable proportions. Then the gaugino masses unify at an intermediate ``mirage'' scale.
This is because the K\"ahler anomalies contributions  \cite{Bagger:1999rd},\cite{Binetruy:2000md}   are suppressed at small $F^T$ and $C=0$ such that  the gaugino mass
splitting at the high energy scale is proportional to the 
beta functions. Since the RG running is governed by the same beta functions, this splitting 
disappears at some intermediate scale. 

The resulting soft terms are 
\begin{eqnarray}
 M_a &=&   {F^T \over T_0 + \overline{T}_0} + {\rm anomaly}
 \;, \nonumber\\
 m_\alpha^2 &=&   \Delta_\alpha  +
(n_\alpha + 3\xi_\alpha) {\vert F^T \vert^2 \over (T_0 + \overline{T}_0)^2 } 
  +  {\rm anomaly}  \;, \nonumber\\
 A_{\alpha\beta\gamma}&=& -{F^T \over T_0 + \overline{T}_0} \left[
3+ n_\alpha + n_\beta + n_\gamma \right] +  {\rm anomaly} \;,
\end{eqnarray}
where $\Delta_\alpha \equiv (1-3 \xi_\alpha) m_{3/2}^2$ and  
the ``anomaly'' contribution to the scalar masses
subsumes possible 1--loop contributions \cite{Binetruy:2000md} as well as
a mixed modulus--anomaly  and 2--loop contributions. 
In the case $\xi_\alpha \sim 1/3$,  the scalar masses
are also suppressed resulting  in the hierarchy 
\begin{equation}
m_{\rm soft} ~\ll~ m_{3/2} ~~, 
\end{equation}
and $m_T \gg m_C \sim m_{3/2}$. Heavy gravitinos and moduli ($\geq 30$ TeV)   
are desirable from the
cosmological perspective since they decay before the nucleosynthesis  
and do not affect the abundances of light elements \cite{Choi:2005uz}. 
In addition, 
this scheme avoids the problem with 
overproduction of gravitinos by heavy moduli \cite{Endo:2006zj}.
The reason is that, at late times, 
the energy density of the Universe is dominated by the $C$ field
which has a mass $\sim m_{3/2}$. Therefore, the branching ratio
for the $C$ decays into gravitinos is suppressed and the ``moduli--induced''
gravitino problem  \cite{Endo:2006zj}  is absent.

\section{Conclusions}

Uplifting AdS vacua in string theory has been  a difficult issue.
One of the popular proposals consistent with spontaneous SUSY breaking is to use
the supersymmetric D--terms.  This requires proper consideration
of the effects due to charged matter which complicates the analysis.

In this work,  we have taken an alternative route. Since one has to include matter effects
anyway, one may as well consider the possibility that  dS and Minkowski vacua arise
due to superpotential interactions involving  hidden matter. 
In this paper, 
we have identified  the local superpotential
structures realizing this situation and studied the resulting SUSY breaking.

We find that, within this scheme, 
the SUSY breaking scale can be explained by the scale of gaugino condensation.
We also find that, when the T--modulus is heavy,  
interesting patterns  of the soft terms  occur. In particular, 
 a matter--dominated SUSY breaking scenario arises. It provides a 
``regularization'' of the traditional 
anomaly mediation scheme as it has most features of the latter
while avoiding tachyonic sfermions. 
Finally, we have shown how mirage mediation is realized  
in the context of spontaneously broken supergravity.\\[0.1cm]

\noindent
{\bf Acknowledgements.} We would like to thank
Kiwoon Choi for valuable discussions.
This work was partially supported by the European Union 6th Framework Program
MRTN-CT-2004-503369 `Quest for Unification' and MRTN-CT-2004-005104
`ForcesUniverse'.

\end{document}